\documentstyle[12pt]{article}
\input epsf
\begin{document}

\begin{titlepage}

\noindent Stockholm\\
USITP 99-4\\
June 1999

\vspace{2cm}

\begin{center}

{\Large DE SITTER SPACE AND SPATIAL TOPOLOGY}

\vspace{2cm}

{\large Ingemar Bengtsson}\footnote{Email address: ingemar@physto.se. 
Supported by NFR.}

\vspace{8mm}

{\large S\"oren Holst}\footnote{Email address: holst@physto.se}

\vspace{1cm}

{\sl Fysikum\\
Stockholm University\\
Box 6730, S-113 85 Stockholm, Sweden}

\vspace{15mm}

{\bf Abstract}

\end{center}

\vspace{1cm}

\noindent Morrow-Jones and Witt have shown that generic spatial topologies 
admit initial data that evolve to locally de Sitter spacetimes under 
Einstein's equations. We simplify their arguments, make them 
a little more general, and solve for the global time evolution of 
the wormhole initial data considered by them. Finally we give explicit 
examples of locally de Sitter domains of development whose 
universal covers cannot be embedded in de Sitter space. 

\end{titlepage}

\noindent {\bf 1. INTRODUCTION.}

\vspace{5mm}

\noindent In a paper entitled "Inflationary data for generic spatial 
topology" Morrow-Jones and Witt \cite{Witt} were able to show that it 
is possible to set initial data on three-spaces of generic topology which 
evolve to locally de Sitter spacetimes under the equation

\begin{equation} R_{{\alpha}{\beta}} = {\lambda}g_{{\alpha}{\beta}} \ , 
\hspace{1cm} {\lambda} > 0 \ . \end{equation}

\noindent During inflation---if it occurs---this is indeed how the Universe 
evolves. Given the continuing interest in both non-trivial spatial 
topologies and inflation it seems worthwhile to explore these results 
a little further. Let us briefly summarize the main points of Morrow-Jones 
and Witt: According to a widely believed conjecture 
\cite{Thurston}, all three-manifolds can be "glued together" from certain 
basic building blocks. These building blocks are Thurston's model 
geometries. There are altogether eight model geometries of which four are 
locally spherically symmetric, namely hyperbolic space ${\bf H}^3$, flat 
space ${\bf E}^3$, 
the three-sphere ${\bf S}^3$, and the handle ${\bf S}^2\times {\bf E}$. 
Naturally we must allow the topologies that can be obtained by taking 
the quotient of a building block with some discrete isometry group 
${\Gamma}$. In a well defined sense "generic" spatial topologies are 
necessarily of the form ${\bf H}^3/{\Gamma}$ \cite{Thurston}, so including 
further model geometries makes the topology generic by default. The first 
observation in ref. \cite{Witt} is that all the four model geometries 
that we have mentioned can be embedded as spatial hypersurfaces in de 
Sitter space, so 
that locally de Sitter initial data can be set on all of them as well as on 
their quotients. Next the "gluing together" is done by taking a connected 
sum, which means that one removes a ball from each piece and identifies 
the resulting boundaries. While it is clear that this can be done in a way 
that preserves the local spherical symmetry of the three-metrics, it is 
not obvious that the full set of initial data for Einstein's equations can 
be chosen so that spacetime remains locally de Sitter. Indeed since there 
are elliptic constraint equations one expects the presence of a handle 
on a sphere (say) 
to affect the geometry all over the sphere, so that attaching a second handle 
might prove difficult. Nevertheless Morrow-Jones and Witt were able to 
show that such "Machian" behaviour is in fact absent provided that the 
minimum radius $R$ of the handles is large enough. To show that the connected 
sums can indeed evolve to locally de Sitter spacetimes Morrow-Jones and Witt 
present a slightly complicated argument that relies on a cosmic Birkhoff 
theorem and analyticity in the various charts. 

In this paper we will rederive and extend the above results in two different 
ways: First we rewrite the de Sitter metric by reparametrizing two of the 
coordinates. This brings two arbitrary functions of each coordinate into 
the metric, and we think of these two functions as providing the general 
solution of the two dimensional wave equation. Effectively, this means that 
we can solve the main problem of ref \cite{Witt} by choosing initial data 
for the wave equation rather than for Einstein's equation. (This method 
can be applied in more general situations but turns out to be useful 
here only because the de Sitter metric is so simple.) Then we present 
a pictorial description of de Sitter space that allows us to drop the 
restriction of local spherical symmetry everywhere, and to see the global 
time evolution of the data at a glance. We use the pictorial 
method to address an interesting point raised by Morrow-Jones and Witt 
(and known to mathematicians \cite{Mess}), namely the existence of smooth 
locally de Sitter domains of development whose universal covers cannot 
be embedded in de Sitter space. It turns out that ${\bf H}^2\times {\bf E}$, 
the fifth of Thurston's model geometries, serves as an example. (On the 
other hand we will find that the example proposed by Morrow-Jones and Witt 
does not work.) 

Throughout, a "spacetime" means the domain of development of some 
complete smooth initial data surface. It will become clear that most of 
the spacetimes constructed are complete only in one time direction, chosen 
to be the future. If we evolve back into the past we will typically encounter 
either Cauchy horizons, Misner type singularities, or both. 
In some cases the solution can be analytically extended across the horizon, 
and then the spatial topology may be revealed as being in some sense not 
what we thought it was. An example would be an ${\bf H}^3$ initial data 
surface, for which the solution can be extended to ordinary de Sitter 
space whose topology is in fact ${\bf S}^3 \times {\bf R}$. In other cases 
there are singularities in the past, perhaps making it impossible to 
continue the solution across the Cauchy horizon. An example would be 
${\bf H}^3/{\Gamma}$ for a suitably chosen discrete group \cite{Koike}. 
We choose to ignore this point, which means that we have nothing to say 
about the issue of how inflation started.

The organization of our paper is as follows: In section 2 we rederive the 
central result of ref. \cite{Witt} by writing the de Sitter metric in a 
form where picking a locally spherically symmetric geometry on an initial 
data slice is achieved by choosing initial data for the two dimensional 
wave equation. In section 3 we show---using the hyperboloid model and 
following Schr\"odinger \cite{Erwin}---how some of Thurston's model geometries 
can be embedded in a de Sitter space. In section 4 we explain how de Sitter 
space can be visualized in intrinsic terms; from here on the argument 
relies to a large extent on pictures. Section 5 is devoted to some of the 
wormhole topologies discussed by Morrow-Jones and Witt \cite{Witt}. Finally, 
in section 6 we construct examples of locally de Sitter spacetimes whose 
universal covers cannot be found as a subset of de Sitter space and in section 
7 we summarize our conclusions.

\vspace{1cm}

\noindent {\bf 2. A USEFUL FORM OF THE DE SITTER METRIC.}

\vspace{5mm}

\noindent We define a locally spherically symmetric three-space as a space 
that can be covered with charts in which the metric takes the form 

\begin{equation} dl^2 = a^2(dr^2 + f^2(r)d{\Omega}^2) \ , \label{3metrik} 
\end{equation}

\noindent where $d{\Omega}^2$ is the metric on the two-sphere ${\bf S}^2$ 
and $a$ is some constant. Examples include four of the five model geometries 
mentioned above as well as their connected sums \cite{Witt}. The point is that 
the connected sum can be taken by removing a sphere symmetrically placed 
within a chart of this type. We then choose an $f$ that interpolates between 
what we need for the model geometries that are to be connected together, 
and we are done. The question to be investigated is therefore: What restrictions 
on $f$ are imposed by the requirement that such a three-space can be 
embedded in a locally de Sitter space?

Our strategy in this section is to reparametrize the de Sitter metric in 
such a way that it contains a function $F(t,r)$; the metric will remain de 
Sitter if and only if $F$ is a solution of the two dimensional wave equation. 
By inspection we will then be able to 
see that choosing the function $f(r)$ and the constant $a$ in the induced 
3-metric at constant $t$ corresponds to choosing initial data for the 
wave equation. A restriction on possible forms of $f$ will arise from the 
demand that the 3-metric shall have Euclidean signature. This then will be 
our answer to the question just asked. 

We begin with two standard expressions for the de Sitter metric, the static form 

\begin{equation} ds^2 = \frac{3}{\lambda}\left( - (1-R^2)dT^2 + 
\frac{dR^2}{1-R^2} + R^2d{\Omega}^2\right) \hspace{1cm}  R < 1  \end{equation}

\noindent and the Kantowski-Sachs form 

\begin{equation} ds^2 = \frac{3}{\lambda}\left( - \frac{dT^2}{T^2 - 1} + 
(T^2 - 1)dR^2 + T^2d{\Omega}^2\right) \hspace{1cm} T > 1 \end{equation}

\noindent where in both cases $d{\Omega}^2$ is the metric on ${\bf S}^2$. 
Neither metric covers the entire spacetime, but they are locally de Sitter 
and this is all we need for our present purposes. 

The static form of the metric can be taken to the conformal gauge 
by means of a rescaling of $R$:

\begin{equation} {\tau} = T \hspace{1cm} \tanh{\rho} = R \hspace{5mm} \Rightarrow 
\hspace{5cm} \end{equation}

\begin{equation} ds^2 = \frac{3}{\lambda}\left( (1 - \tanh^2{\rho})(-d{\tau}^2 + 
d{\rho}^2) + \tanh^2{\rho}d{\Omega}^2 \right) \ . \end{equation}

\noindent Define 

\begin{equation} U = {\tau} - {\rho} \hspace{1cm} V = {\tau} + {\rho} 
\end{equation}

\noindent and perform the reparametrization 

\begin{equation} U = U(u) \hspace{1cm} V = V(v) \ . \end{equation}

\noindent The metric then takes the form

\begin{equation} ds^2 = \frac{3}{\lambda}\left( - (1 - \tanh^2{F})
\frac{\partial U}{\partial u}\frac{\partial V}{\partial v}dudv + 
\tanh^2{F}d{\Omega}^2\right) \end{equation}

\noindent where 

\begin{equation} F = \frac{1}{2}(V(v) - U(u)) \ . \end{equation}

\noindent But this means that $F$ is an arbitrary solution of the two 
dimensional wave equation. Transforming to a new set of coordinates 

\begin{equation} u = t - r \hspace{1cm} v = t + r \end{equation}

\noindent the metric takes the form 

\begin{equation} ds^2 = \frac{3}{\lambda}\left( (1 - \tanh^2{F})
(F^{\prime 2} - \dot{F}^2)(-dt^2 + dr^2) + 
\tanh^2{F}d{\Omega}^2\right) \ , \end{equation}

\noindent where $F$ is an arbitrary solution of the two dimensional wave 
equation in the coordinates $r$ and $t$, a prime denotes differentiation 
with respect to $r$ and a dot with respect to $t$. This is the form of 
the metric that we were after. Treating the 
Kantowski-Sachs metric in the same way we obtain 

\begin{equation} ds^2 = \frac{3}{\lambda}\left( (\coth^2{F} - 1)
(\dot{F}^2 - F^{\prime 2})(-dt^2 + dr^2) + 
\coth^2{F}d{\Omega}^2\right) \ . \label{4metrik} \end{equation}

\noindent For our purposes this is all we need. 

To embed a locally spherically symmetric three-space with the metric 
(\ref{3metrik}) all that is needed is to choose initial data for the 
two dimensional wave equation. The initial values of $F$ and $\dot{F}$ 
at any fixed value of $t = t_0$ are at our disposal, so if $|af| > 
\sqrt{3/{\lambda}}$ we set 

\begin{equation} g_{{\theta}{\theta}} = \frac{3}{\lambda}\coth^2{F} = 
a^2f^2 \end{equation}

\begin{equation} g_{rr} = \frac{3}{\lambda}(\coth^2{F} - 1)
(\dot{F}^2 - F^{\prime 2}) = a^2 \ . \end{equation}

\noindent and we are done: The spatial metric at $t = t_0$ is the 
locally spherically symmetric metric (\ref{3metrik}), and the initial 
values determine a solution $F(r,t)$ of the wave equation that yields the 
locally de Sitter metric (\ref{4metrik}). There are no additional restrictions 
on the function $f$ in this case. 

The story is different if $|af| < \sqrt{3/{\lambda}}$. This time we set 

\begin{equation} g_{{\theta}{\theta}} = \frac{3}{\lambda}\tanh^2{F} = 
a^2f^2 \end{equation}

\begin{equation} g_{rr} = \frac{3}{\lambda}(1 - \tanh^2{F})
(F^{\prime 2} - \dot{F}^2) = a^2 \ . \end{equation}

\noindent Now there is an additional restriction since $r$ will be a spatial 
coordinate only if

\begin{equation} g_{rr} > 0 \hspace{5mm} \Rightarrow \hspace{5mm} 
F^{\prime 2} - \dot{F}^2 > 0 \ . \label{18} \end{equation}

\noindent Hence $\dot{F}$ cannot be chosen quite arbitrarily. Moreover from 
the relation between $f$ and $F$ it follows that  

\begin{equation} f^{\prime} = 0 \hspace{5mm} \Rightarrow \hspace{5mm} 
F^{\prime} = 0 \ . \label{19} \end{equation}

\noindent Eqs. (\ref{18}) and (\ref{19}) imply that we cannot allow 
the function $f$ in the three-metric 
(\ref{3metrik}) to attain a minimum or a maximum at a radius smaller 
than $\sqrt{3/{\lambda}}$. In particular only handles whose 
minimum radius exceeds $\sqrt{3/{\lambda}}$ can occur. This completes our 
rederivation of the central result of ref. \cite{Witt}.

The reader may worry that $|af|$ is a function, and it may be larger or 
smaller than $\sqrt{3/{\lambda}}$ depending on its argument. Actually 
there is no problem here, as an explicit example will make clear. Set 

\begin{equation} F = \frac{1}{2}\ln{\frac{r-t}{r+t}} \ , \end{equation} 

\noindent which is a real solution of the wave equation if $r > t$. 
It is easy to show that 

\begin{equation} \coth{F} = - \frac{r}{t} \end{equation}

\noindent and that the locally de Sitter metric (\ref{4metrik}) becomes 

\begin{equation} ds^2 = \frac{3}{\lambda}\frac{1}{t^2}( - dt^2 + 
dr^2 + r^2d{\Omega}^2) \ , \end{equation}

\noindent which represents a portion of de Sitter space foliated by 
flat spaces. When $r < t$ it happens that the argument of the 
logarithm becomes negative so that $F$ develops an imaginary part. 
This is just what is needed in order to turn the cotangens hyperbolicus 
into tangens hyperbolicus, so that the other form of the metric applies.

Finally we observe that our argument can be varied a little. An 
alternative form of the intrinsic de Sitter metric is 

\begin{equation} ds^2 = \frac{3}{\lambda}\left( - \frac{dT^2}{T^2 + 1} 
+ (T^2 + 1)dR^2 + T^2d{\sigma}^2\right) \end{equation} 

\noindent where $d{\sigma}^2$ is the metric on ${\bf H}^2$. Treating this 
line element in the same way as above we arrive at 

\begin{equation} ds^2 = \frac{3}{\lambda}\left( (1 + \tan^2{F})
(\dot{F}^2 - F^{\prime 2})(- dt^2 + dr^2) + \tan^2{F}d{\sigma}^2
\right) \ . \label{21} \end{equation}

\noindent Possible initial data slices now include hyperbolic three-space 
${\bf H}^3$ and the hyperbolic handle ${\bf H}^2\times {\bf E}$. 
\vspace{1cm}

\noindent {\bf 3. MODEL GEOMETRIES.}

\vspace{5mm}

\noindent This section is intended to remind the reader of some 
well known facts \cite{Erwin}. If (for convenience) we set ${\lambda} = 3$ 
then de Sitter space can be defined as the hyperboloid 

\begin{equation} X^2 + Y^2 + Z^2 + U^2 - V^2 = 1 \end{equation}

\noindent sitting inside a five dimensional Minkowski space with the metric 

\begin{equation} ds^2 = dX^2 + dY^2 + dZ^2 + dU^2 - dV^2 \ . \end{equation}

\noindent The isometry group of de Sitter space is therefore the Lorentz 
group $SO(4,1)$. The embedding space coordinates are very useful for 
many calculations. In particular, let us consider various spacelike 
hypersurfaces in de Sitter space. 

Intersecting the hyperboloid with the spacelike hyperplanes 

\begin{equation} V = \sinh{t} \hspace{5mm} \Rightarrow \hspace{5mm} X^2 + Y^2 
+ Z^2 + U^2 = \cosh^2{t} \end{equation}

\noindent we obtain a foliation of de Sitter space with 3-spheres. These 
3-spheres turn out to be totally umbilic (that is to say their first and second 
fundamental forms are proportional). They are totally geodesic and have 
unit radius if and only if the spacelike plane goes through the origin. 
Moving the family of spacelike planes around with isometries it is clear 
that there is a totally geodesic sphere through any point in de Sitter space. 

Intersecting the hyperboloid with the timelike hyperplanes 

\begin{equation} U = \cosh{t} \end{equation}

\noindent we get a foliation of a region of de Sitter space with totally 
umbilic hyperbolic 3-spaces. (Intersecting with a timelike hyperplane through 
the origin yields a totally geodesic three-dimensional de Sitter space.) 
Intersecting with null hyperplanes 

\begin{equation} U + V = \frac{1}{t} \ , \hspace{8mm} t > 0 \end{equation}

\noindent we obtain a foliation of "one half" of de Sitter space with totally 
umbilic flat 3-spaces. Intersection with a null plane through the origin 
yields a totally geodesic null hypersurface. Two null planes in five dimensional 
Minkowski space intersect in a 3-space (unless the null planes are parallel), 
and the intersection of this 3-space with the de Sitter hyperboloid is a 
2-sphere.

We have seen how three of the model geometries (${\bf S}^3$, ${\bf H}^3$ and 
${\bf E}^3$) can be embedded in de Sitter space. To see that the handle can 
occur, intersect with the upper branch of the spacelike surface 

\begin{equation} V^2 - U^2 = \sinh^2{t} \hspace{5mm} \Rightarrow \hspace{5mm} 
X^2 + Y^2 + Z^2 = \cosh^2{t} \ . \end{equation}

\noindent This is indeed the handle ${\bf S}^2\times {\bf E}$, with the radius 
of the 2-sphere necessarily larger than one. Finally we can intersect the 
hyperboloid with the surface 

\begin{equation} Z^2 + U^2 = \cosh^2{t} \hspace{5mm} \Rightarrow \hspace{5mm} 
X^2 + Y^2 - V^2 = - \sinh^2{t} \ . \end{equation} 

\noindent The intersection has the geometry of ${\bf H}^2\times {\bf S}^1$. 
Now ${\bf H}^2\times {\bf E}$ is also one of Thurston's model geometries. 
If we can "unwrap" the circle it follows that this model geometry also can 
occur in a locally de Sitter spacetime. Actually the function $F$ in the 
metric (\ref{21}) can be arranged so that this is true but it appears that 
this spatial geometry cannot be embedded in de Sitter space. We will return 
to this point in section 6.

Our last two examples are not umbilic surfaces but they do inherit enough 
Killing vectors from de Sitter space to be homogeneous spaces, and their 
second fundamental forms share the symmetries of the induced metric.

\vspace{1cm}

\noindent {\bf 4. VISUALIZING DE SITTER SPACE.}

\vspace{5mm}

\noindent We would now like to show some pictures of the above. This can 
be done if we restrict ourselves to 2+1 dimensions and use the methods of 
ref. \cite{adS}, which have proved quite useful to derive various properties 
of locally anti-de Sitter spaces. Naturally there are features of 3+1 
dimensional spacetimes that cannot be seen in the lower dimension, but 
as far as the properties of locally spherically symmetric spacetimes are 
concerned the restriction is harmless. Thus we consider the hyperboloid 

\begin{equation} X^2 + Y^2 + U^2 - V^2 = 1 \end{equation}

\noindent in ordinary Minkowski space. We use a function of $V$ as 
an intrinsic time coordinate; at constant $V$ space is a sphere. As 
map-makers know we cannot make a two dimensional map of a sphere unless we 
give up manifest spherical symmetry; our choice is to make a stereographic 
projection onto the unit disk of each hemisphere separately. Then we 
arrive at the coordinate system $(x,y,t)$, where 

\begin{equation} X = \frac{2x}{1+r^2}\frac{1}{\cos{t}} \hspace{2cm} 
Y = \frac{2y}{1+r^2}\frac{1}{\cos{t}} \end{equation}

\begin{equation} U = \pm \frac{1-r^2}{1+r^2}\frac{1}{\cos{t}} \hspace{18mm} 
V = \tan{t} \ , \end{equation}

\noindent and we use the upper sign in the Northern hemisphere and the 
lower sign in the Southern. Note that 

\begin{equation} r^2 \equiv x^2 + y^2 \leq 1 \ . \end{equation}

\noindent The metric in these coordinates is 

\begin{equation} ds^2 = \frac{1}{\cos^2{t}}\left( -dt^2 + \frac{4}{(1+r^2)^2}
(dx^2 + dy^2)\right) \ . \end{equation}

\noindent The time coordinate $t$ goes from $- {\pi}/2$ to ${\pi}/2$, and 
we can attach a conformal boundary consisting of two spheres at $t = \pm 
{\pi}/2$. We can now draw pictures in coordinate space. Our first picture 
shows de Sitter space as two beer cans whose cylindrical boundaries are to 
be identified. The scaling of the radial coordinate was chosen so that the 
slope of a light ray is a function of $r$ only. 

We observe that all timelike and null geodesics acquire a future endpoint 
on future infinity $i^+$ and a past endpoint on past infinity $i^-$, while 
all spatial geodesics are closed and have circumference $2{\pi}$.

Totally geodesic surfaces are easy to find since they are the intersection 
of the hyperboloid with hyperplanes through the origin in embedding space. 
Let us begin with totally geodesic null surfaces which result if the 
hyperplane through the origin is chosen to be null. It is easy to check 
that there is a one-to-one correspondence between such surfaces and the 
set of lightcones with a vertex on $i^+$. (For the twistor theorist this 
means that $i^+$ can be thought of as mini-twistor space.) The next picture 
shows two such surfaces, chosen so that they are rotationally symmetric 
in our picture. 

\begin{figure}[t]
  \hspace*{5mm}
  \begin{minipage}[c]{125mm}
  \hspace*{-8mm}
  \setlength{\epsfxsize}{140mm}
  \epsfbox{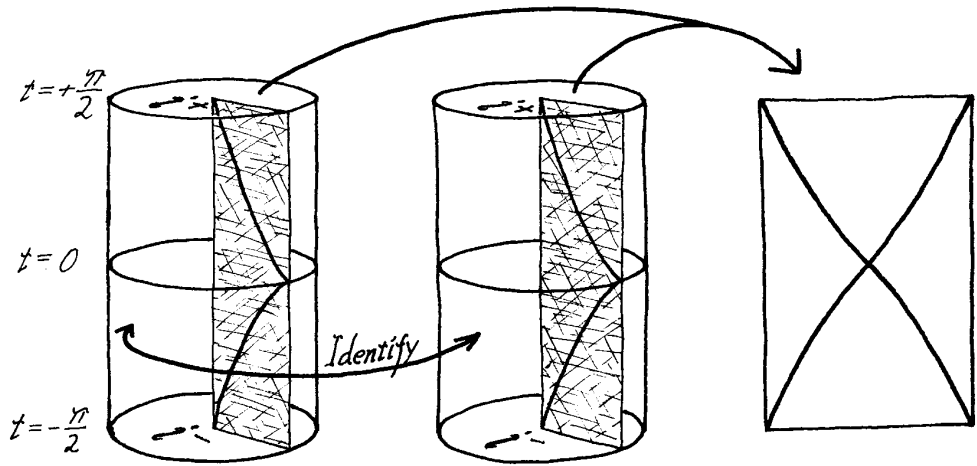}
  \vspace*{-7mm}
  \caption{Two beer cans representing de Sitter space. 
Surfaces of constant $t$ are pairs of disks representing the two hemispheres 
of a stereographically projected 2-sphere. The pair of disks on top of the 
cans represents $i^+$. The shaded surface is a Carter-Penrose diagram, each 
interior point of which represents a circle. The curves on this surface are 
light rays. Such diagrams are usually drawn so that the slope of a radial 
light ray is unity but this is not the case here. Our radial coordinate is 
chosen so that the slope of any light ray depends on $r$ only.}
\end{minipage} 
\hfil
\end{figure}

The isometry group of 2+1 dimensional de Sitter space is the Lorentz 
group $SO(3,1)$, which acts 
like the group of M\"obius transformations on $i^+$. In particular a Lorentz 
boost $J_{UV}$ has a bifurcate Killing horizon that coincides with the light 
cones with vertices at $r = 0$; the flow is timelike inside the lightcones 
and spacelike outside. By symmetry all light cones with vertices on $i^+$ 
are Killing horizons. Two important properties follow immediately: First 
all spacelike loops going around such a cone have equal length (equal 
to $2{\pi}$ with our choice of ${\lambda}$). As a consequence the intersection 
of two light cones with vertices on $i^+$ is always a circle with 
circumference $2{\pi}$. 

\begin{figure}[t]
  \hspace*{10mm}
  \begin{minipage}[c]{115mm}
  \hspace*{-5mm}
  \setlength{\epsfxsize}{110mm}
  \epsfbox{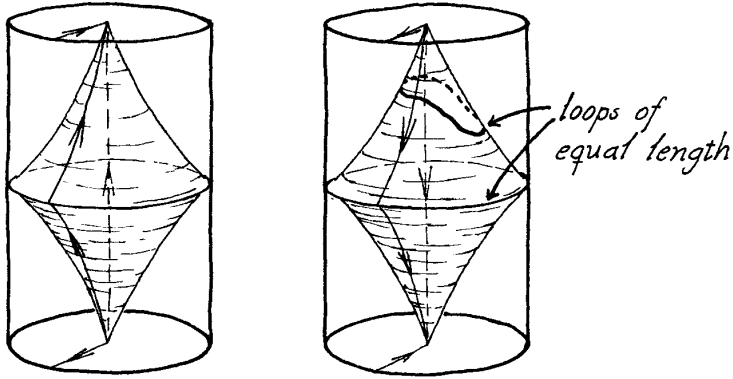}
  \caption{A lightcone with its vertex on $i^+$ is always totally 
geodesic. A pair of such lightcones make up a bifurcate Killing horizon; 
the flow of the relevant Killing vector $J_{UV}$ is indicated by the arrows. 
Because the surface is a 
Killing horizon all spacelike loops going around it have equal length. This 
helps to give a feeling for how distances are distorted by the picture.}
\end{minipage} 
\hfil
\end{figure}

Lorentz boosts correspond to hyperbolic M\"obius transformations on
$i^+$ while rotations correspond to elliptic M\"obius transformations;
rotations have two timelike lines of fixed points which can be made
into world lines of point particles, if this is wanted \cite{Deser}. A
general Lorentz transformation can be written as a combination of a
boost and a rotation and is called a four screw. It has no fixed point
inside de Sitter space but two fixed points on $i^+$, where it gives
what one calls a loxodromic M\"obius transformation.  Finally a null
boost has a single fixed point on $i^+$ and a single light cone for
its (degenerate) Killing horizon; there are two lightlike lines of
fixed points going through de Sitter space.

We may note that if we identify points along the Killing flow lines of a 
boost (say $J_{UV}$ for definiteness) 
we obtain---in the regions where the flow is spacelike---the spatial 
topology of a torus. This is the de Sitter analogue of Misner space; the 
anti-de Sitter analogue is the BTZ black hole \cite{adS}.

Let us now consider foliations by spacelike surfaces. In particular we want 
to draw pictures of the two dimensional analogues ${\bf S}^2$, ${\bf E}^2$ and 
${\bf S}^1\times {\bf E}$ of the three dimensional model geometries discussed 
in section 3. Note that the distinct three dimensional possibilities 
${\bf H}^2\times {\bf S}^1$ and ${\bf S}^2\times {\bf E}$ collapse to one 
example only in 2+1 dimensions. We begin with spheres. A foliation with 
surfaces of constant 
$t$ consists of contracting and expanding totally umbilic 2-spheres; only the 
pair of disks at $t = 0$ represents a totally geodesic sphere. Moving this 
exceptional sphere around with suitable isometries will give a foliation 
with totally geodesic spheres of 
the interior of a pair of totally geodesic null surfaces. 

\begin{figure}[h]
  \hspace*{3mm}
  \begin{minipage}[c]{130mm}
  \hspace*{-5mm}
  \setlength{\epsfxsize}{140mm}
  \epsfbox{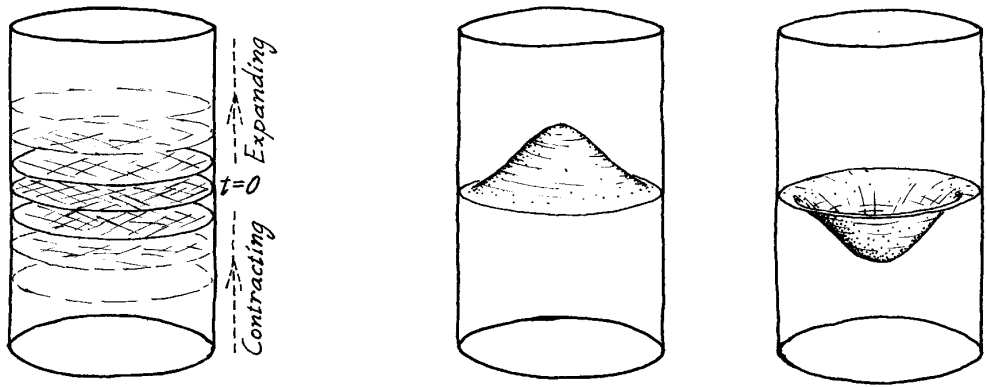}
  \vspace*{-5mm}
  \caption{On the left de Sitter space is shown foliated by 
expanding and contracting spheres; only the exceptional sphere at $t=0$ 
is totally geodesic (only one can is shown). To the right the totally 
geodesic sphere has been moved to another position by means of an 
isometry.}
\end{minipage} 
\end{figure}
 
\begin{figure}[!h]
  \setlength{\epsfxsize}{140mm}
  \epsfbox{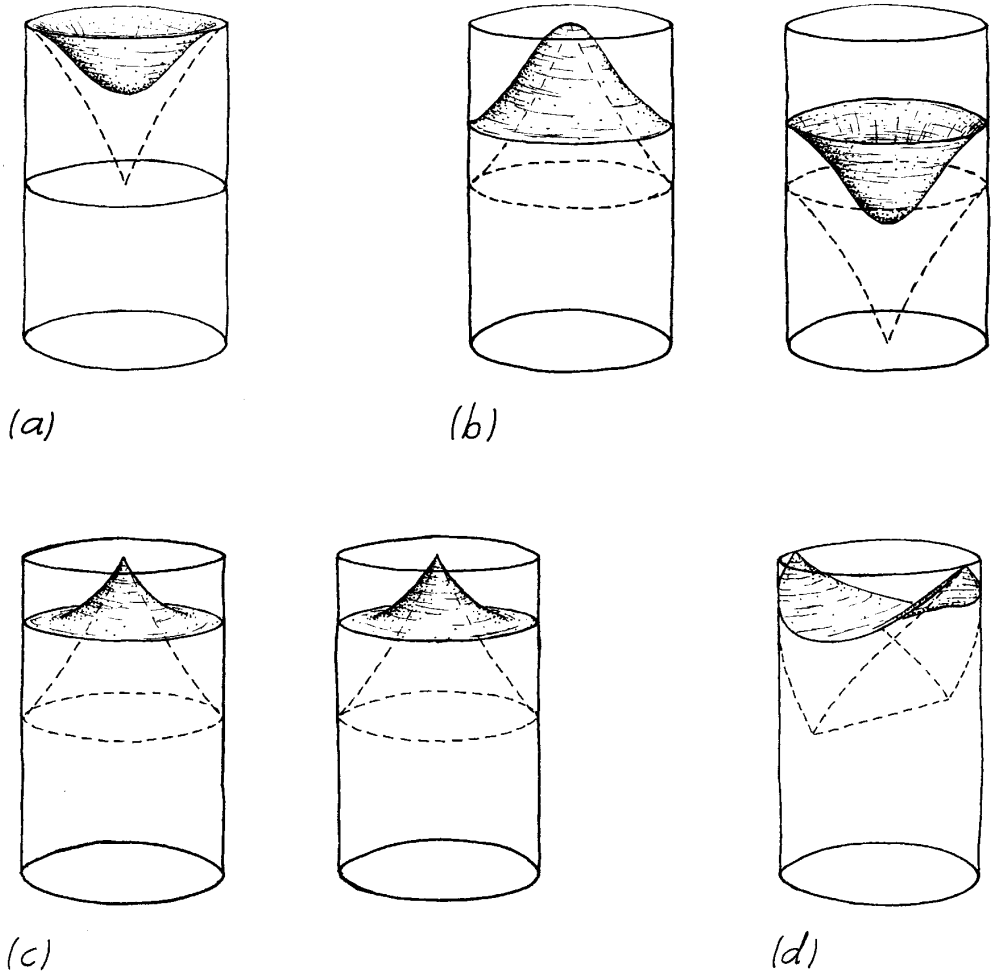}
  \caption{\noindent Other model geometries, with their Cauchy horizons 
    shown dashed: 
(a) A hyperbolic plane ${\bf H}^2$ (only one can is shown---the other 
is empty). 
(b) A flat space ${\bf E}^2$. 
(c) A cylinder ${\bf S}^1\times {\bf E}$, placed so that its 
symmetry is manifest in the picture. 
(d) The same cylinder moved to a new position by an isometry 
(only one can is shown---the other is identical).}
\vspace*{10mm}
\end{figure}

We will take special interest in surfaces that behave asymptotically
like one of the model geometries ${\bf H}^2$, ${\bf E}^2$ and ${\bf
  S}^1\times {\bf E}$. This means that they have to intersect $i^+$ in
a circle, a point or two points, respectively (Figure 4). As we saw in
section 3, intersecting the de Sitter hyperboloid with a family of
null planes gives a foliation in terms of flat surfaces, while
intersecting with timelike planes sufficiently far from the origin
gives a foliation with expanding hyperbolic planes. Following section
3 we can also find a foliation with cylinders; we draw this in two
different ways to remind the reader that if a given spatial geometry
is displaced by an isometry then its appearance in our pictures will
change. In all three cases the foliation covers only a region of de
Sitter space. The boundary of this region is the Cauchy horizon, which
is a null surface. In the case of the handle the Cauchy horizon grows
from a spacelike geodesic which is a circle and has circumference
$2{\pi}$.

\vspace{1cm}

\noindent {\bf 5. SPACES WITH ASYMPTOTIC REGIONS.}

\vspace{5mm}

\noindent Given the pictures of the various model geometries it is 
straightforward to draw their connected sums. In fact we do not have 
to insist that the surfaces be everywhere locally spherically 
symmetric. We can wiggle them this way and that. As an example, let us 
(with Morrow-Jones and Witt \cite{Witt}) consider the problem of attaching 
asymptotic regions to some spatial slice. Then the thing to 
remember is that a hyperbolic plane always intersects $i^+$ in a 
circle, while flat planes and handles intersect $i^+$ in respectively one 
and two points. In our picture what we have to do is to deform the sphere 
so that it touches $i^+$ in the "right" way (Figure 5). 

\begin{figure}[!b]
  \hspace*{3mm}
  \begin{minipage}[c]{130mm}
  \hspace*{-7mm}
  \setlength{\epsfxsize}{145mm}
  \epsfbox{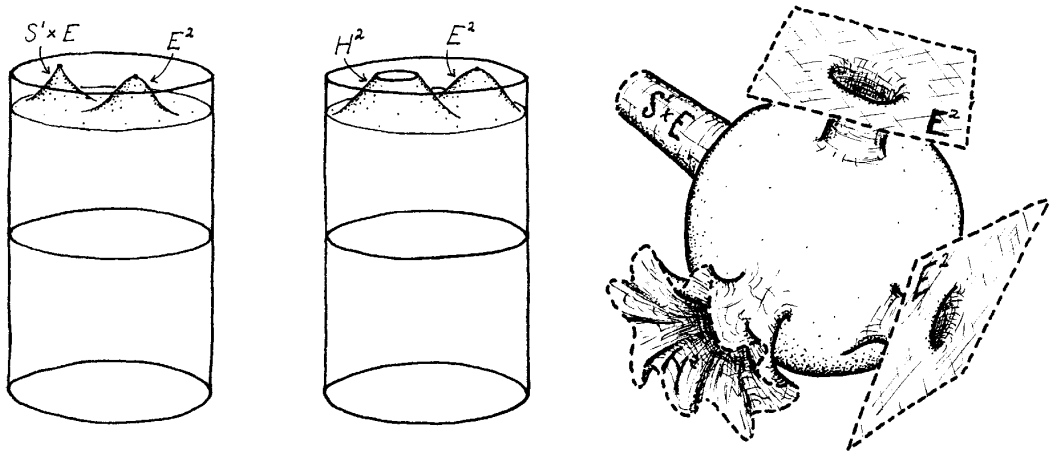}
  \vspace*{-2mm}
  \caption{A two-sphere with a number of asymptotic regions attached. 
Two of the asymptotic regions meet $i^+$ in such a way that they are flat, 
one is a cylinder, and one a hyperbolic plane (and the latter contains so 
much space that it is difficult to draw it in ordinary 3-space).}
\end{minipage} 
\end{figure}

It is interesting to ask for the minimum size of the disk that must be 
removed from the sphere when an asymptotic region is attached. Consider a 
flat asymptotic region, say. Then we must certainly remove that portion of 
the sphere that lies in the causal past of the point where the flat space 
touches $i^+$. This is a disk whose circumference is a loop surrounding 
a light cone with its vertex on $i^+$, and as we have seen it necessarily 
has circumference $2{\pi}$. But we can place the asymptotic region as close 
to this light cone as we want, and therefore the circumference of the disk 
that we must remove is bounded from below by $2{\pi}$. We get the same bound 
if the asymptotic region is a cylinder or a hyperbolic plane. 

Among the 3-spaces considered by Morrow-Jones and Witt \cite{Witt} is a 
set of three asymptotically flat spaces connected by two handles. They 
make the claim that although this space will (given a suitable form 
of its extrinsic curvature) evolve to a locally de Sitter spacetime, the 
universal cover of the domain of development cannot be embedded in de 
Sitter space. Actually this is incorrect.

The argument \cite{Witt} goes as follows: In four spacetime 
dimensions the space is simply connected so that we do not have to 
discuss whether some quotient of de Sitter space will serve as embedding 
space---it must be de Sitter space itself. This is indeed so. Morrow-Jones 
and Witt next observe that two flat spaces in de Sitter space necessarily 
intersect in a sphere (unless they occur at different "times"). This is 
also correct---in embedding space it is the statement that two non-parallel 
null planes will intersect in a 3-space, and this 3-space intersects the 
hyperboloid in a sphere. Finally it is concluded 
from the picture of the spatial slice that two of the flat regions do not 
so intersect---but this is not valid since in fact all 
the spatial planes do intersect, although the two at the ends do so (as it 
were) within the disks that were removed in attaching the handles that connect 
them to the plane in the middle. 

\begin{figure}[t]
  \hspace*{2mm}
  \begin{minipage}[c]{130mm}
  \hspace*{-7mm}
  \setlength{\epsfxsize}{140mm}
  \epsfbox{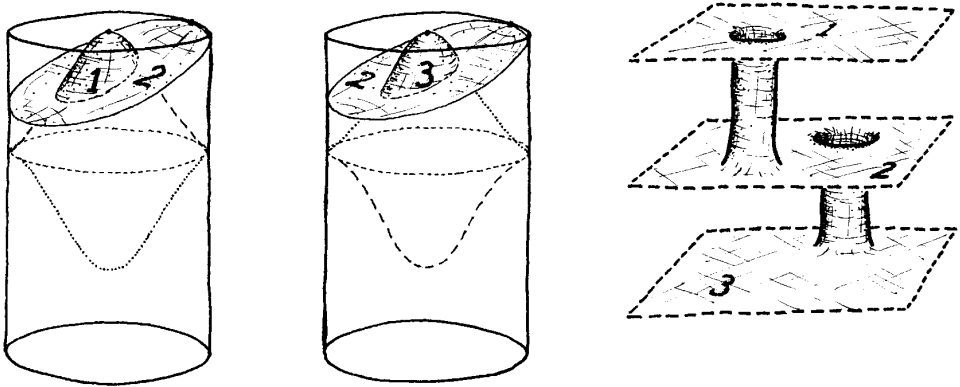}
  \caption{A spatial slice consisting of three flat spaces connected 
by two handles, and its embedding in de Sitter space. Note that had the 
intermediate plane 2 not been there then the planes 1 and 3 would have been 
continued downwards in the cans as indicated by the dashed and dotted lines, 
respectively. As a result they would have intersected in a circle. This 
behaviour could not have been guessed just by looking at the schematic 
picture on the right.}
\end{minipage} 
\end{figure}

If this sounds obscure, we hope that it is made clear by a glance at
our two beer cans, embellished with a spatial surface carrying
precisely the geometry considered by Morrow-Jones and Witt (Figure 6).
Although the picture is restricted to 2+1 dimensional de Sitter space
the restriction is irrelevant here. Moreover, in drawing this picture
we have in effect solved for the time evolution of these initial data.
By the way, the restriction on the minimum radius of a connecting
handle (given in ref. \cite{Witt} and in section 2) also follows from
an argument that parallels the argument about the minimum size of the
disk that must be removed when attaching an asymptotic region from the
sphere.

Nevertheless there are locally de Sitter spacetimes whose universal covers 
cannot be regarded as subsets of de Sitter space; we devote section 6 to this 
topic.

\vspace{1cm}

\noindent {\bf 6. WHY DE SITTER SPACE IS NOT ENOUGH.}

\vspace{5mm}

\noindent A relativistic space form, by definition, is a complete Lorentzian 
manifold of constant curvature. They have been completely classified 
\cite{Calabi}. The universal cover of a relativistic space form is de Sitter 
space, anti-de Sitter space or Minkowski space depending on the curvature. 
Our definition of a spacetime was made in the spirit of canonical gravity 
and does not require completeness. All that we require is the domain of 
development of some smooth initial data surface. The classification of such 
spacetimes is a much harder problem and---at least to our knowledge---it 
has been achieved only in 2+1 dimensions \cite{Mess}. In this connection 
it was pointed out by Mess \cite{Mess} that there are smooth locally de Sitter 
domains of development (complete to the future) whose universal covers cannot 
be embedded in de Sitter space. 

So how can we obtain a smooth surface carrying locally de Sitter data
that cannot be embedded in de Sitter space? This is not so difficult.
Suppose that we remove a pair of antipodal points from a sphere so
that its topology becomes that of a cylinder, and then go to its
universal covering space. A similar man\oe uvre carried through for
2+1 dimensional de Sitter space entails removing two antipodal
timelike lines and going to the universal covering space of the
remaining (incomplete) spacetime. To draw the picture it is convenient
to place the timelike lines to be removed on the boundary of the cans.
The universal covering space then consists of an infinite set of cans
with lines cut out of the boundary; the sides of the cans are to be
identified pairwise in an obvious manner (Figure 7). Now consider a
particular can and insert a "handle" (actually half a handle) that
touches what used to be $i^+$ in precisely the two points that we have
removed from the top. Draw a similar surface in all the other cans.
This is a smooth spacelike surface of topology ${\bf R}^2$---in
effect, we have "unwrapped" the cylinder. Its intrinsic geometry is
flat but unlike the family of embedded flat spaces considered earlier
it is not an umbilic surface. Moreover its Cauchy development is
smooth and locally de Sitter by construction, and it is complete to
the future. If we try to add a conformal boundary $i^+$ we run into
problems, but then a smooth conformal completion of the domain of
development was never promised. In its past there is a Cauchy horizon
that grows from a non-closed spacelike geodesic of infinite
length---which is one way to see that this spacetime definitely cannot
be embedded in de Sitter space since the latter does not have such
geodesics. Hence this is the example that we were after.

\begin{figure}[h]
  \hspace*{2mm}
  \begin{minipage}[c]{130mm}
  \hspace*{-5mm}
  \setlength{\epsfxsize}{140mm}
  \epsfbox{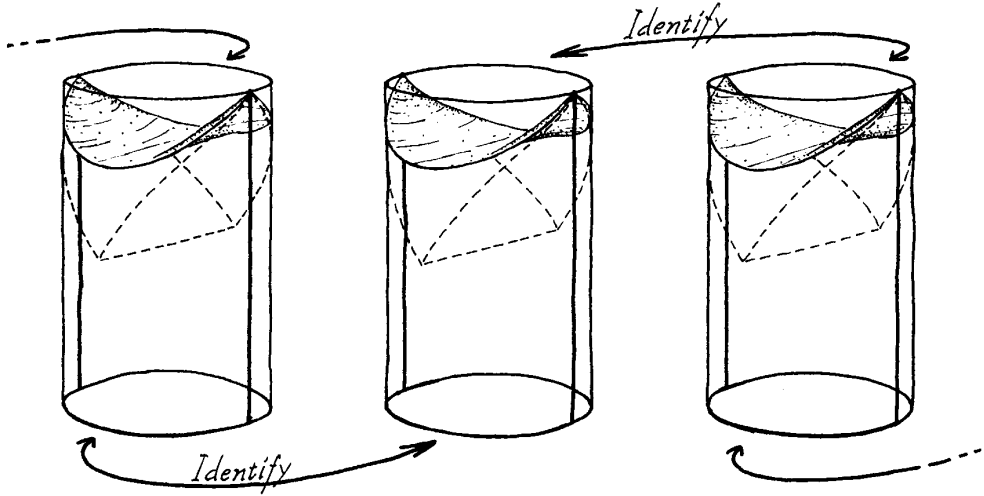}
  \caption{A smooth locally de Sitter and future complete domain 
of development; in effect an "unrolled" handle. Now there is an infinite 
set of cans (only three cans are drawn) with sides identified as shown. 
This is possible because a pair of timelike geodesics on each boundary 
has been removed. The domain of development of the embedded surface is 
simply connected, yet it cannot be embedded in de Sitter space. Note that 
the timelike geodesics that we have removed lie outside the Cauchy 
development (indicated with dashed lines) of the surface.}
\end{minipage} 
\end{figure}

Let us mention in passing that if we start from the de Sitter analogue 
of Misner space (as briefly described in section 4) and then "cut out 
a wedge" bounded by timelike surfaces---that is, if we identify points along 
the flow lines of the Killing vector $J_{XY}$---then the spatial topology 
is again a torus, but a torus whose covering space (in the generic case) 
is the spacetime that we have just discussed.

The example just given works specifically in 2+1 dimensions, but it is 
not difficult to modify it so that it works in 3+1 dimensions. We start 
out by considering the model geometry ${\bf H}^2\times {\bf S}^1$ embedded 
as a hypersurface in de Sitter space through

\begin{equation} Y^2 + U^2 = \cosh^2{\tau} \hspace{5mm} \Leftrightarrow 
\hspace{5mm} X^2 + Z^2 - V^2 = - \sinh^2{\tau} \ . \end{equation} 

\noindent Our aim is to "unwrap" the circle. The Cauchy development of 
the embedded hypersurface is the future of the spatial geodesic 

\begin{equation} X = Z = V = 0 \ . \end{equation} 

\noindent The circle becomes unwrapped if we are able to unwrap the flow 
lines of the Killing vector $J_{YU}$ (that generates rotations). To succeed 
in this we must remove the fixed points of this Killing vector from de 
Sitter space; in equations the fixed points are given by

\begin{equation} J_{YU} = Y\partial_U - U\partial_Y = 0 \hspace{5mm} 
\Rightarrow \hspace{5mm} Y = U = 0 \ . \end{equation}

\noindent Metrically this is a 1+1 dimensional de Sitter space; for us 
it is important to observe that it has zero intersection with the Cauchy 
development of the embedded hypersurface. On the other hand the fixed 
point set and the embedded hypersurface touch $i^+$ in the same circle. 
If we think of de Sitter space as a set of foliating 3-spheres we see 
that the fixed point set gives a circle in each 3-sphere---and a circle 
is precisely what we must remove in order to make a 3-sphere multiply 
connected. Therefore once the fixed point set has been removed we have 
an incomplete, multiply connected spacetime and we can go to its universal 
covering space. Having done this we have an 
embedding of Thurston's model geometry ${\bf H}^2\times {\bf E}$ in an 
incomplete spacetime that is "larger" than de Sitter space; nevertheless 
the Cauchy development of our model geometry is not only simply connected 
but also complete to the future. This Cauchy development is in itself a 
spacetime in our sense. 

To visualize this construction one can draw a series of equal-$t$ slices 
of 3+1 dimensional de Sitter space, that is a series of stereographically 
projected 3-spheres, and study how the hypersurface ${\bf H}^2\times 
{\bf S}^1$ intersects these slices. In effect all that one has to do 
is to take equal-$t$ slices of fig. 7 and rotate them around a suitable 
axis. The details are left to the interested reader.

We have embedded the model geometry ${\bf H}^2\times {\bf E}$ in a locally 
de Sitter spacetime which cannot itself be found as a subset of de Sitter 
space. To see that there is no other way to embed this model geometry in 
de Sitter space we may use eq. (\ref{21}) to prove that its extrinsic 
geometry is determined by the requirement that it is 
embedded in a locally de Sitter space. Therefore its Cauchy development 
has to be precisely the spacetime that we just constructed, and this 
spacetime cannot be a subset of de Sitter space since---like its 
2+1 dimensional counterpart---it grows from an infinitely long
non-closed spacelike geodesic.

\vspace{1cm}

\noindent {\bf 7. CONCLUSIONS.}

\vspace{5mm}

\noindent Our conclusions can be summarized like this: 

One can write the de Sitter metric in a form that allows one to choose 
any allowed locally spherically symmetric geometry on a spacelike slice 
by choosing initial data for the two dimensional wave equation. 

A similar form exists for spatial geometries of the form 
(hyperbolic plane)$\times $ (something). 

Using a pictorial presentation the restriction to local spherical symmetry 
everywhere can be dropped and the Cauchy development of any spatial 
slice is easily studied. 

The model geometry ${\bf H}^2\times {\bf E}$ provides an explicit 
example of a smooth locally de Sitter domain of development 
whose universal cover cannot be embedded in de Sitter space. 

On the other hand three flat asymptotic regions connected by two wormholes 
can be so embedded ({\it pace} previous claims).

As a final comment we observe that de Sitter space is very different from 
the other two relativistic space forms, Minkowski space and anti-de Sitter 
space. In de Sitter space asymptotic regions of spacelike slices occur 
where the slices approach future infinity. Hence the issue of connecting 
different asymptotic regions with causal curves \cite{censur} does not even 
arise. A little thought will also convince the reader that our examples 
of simply connected locally de Sitter spacetimes that cannot be 
embedded in de Sitter space came about precisely because the conformal 
boundary (or more accurately the would be conformal boundary) is a 
spacelike surface. Hence this behaviour cannot occur for locally flat or 
anti-de Sitter spacetimes---a fact which is perfectly well known \cite{Mess}, 
but perhaps we have managed to convey some extra feeling for why it is 
true.

\vspace{2cm}

\noindent \underline{Acknowledgements:} We thank Dieter Brill for suggesting 
this problem, and Stefan \AA minneborg and Greg Galloway for discussions and 
help. IB was supported by the NFR.

\end{document}